\documentclass{PoS}
\newcommand{\be}{\begin{equation}}  
\newcommand{\ee}{\end{equation}}  
\newcommand{\beq}{\begin{eqnarray}}  
\newcommand{\eeq}{\end{eqnarray}}
\newcommand{\Dlr}{\buildrel \leftrightarrow \over D\raise-1pt\hbox{}}

\usepackage{amsmath}
\usepackage{amsfonts,amssymb}
\title{Nucleon observables and axial  charges of other baryons using twisted mass fermions}

\ShortTitle{Nucleon observables and axial charges of other baryons}

\author{\speaker{Constantia  Alexandrou }\\
         Department of Physics, University of Cyprus, P.O. Box 20537, 1678 Nicosia, Cyprus, \\  
 Computation-based Science and Technology Research  
    Center, Cyprus Institute, 20 Kavafi Str., Nicosia 2121, Cyprus, \\
         NIC, DESY, Platanenallee 6, D-15738 Zeuthen, Germany\\
        E-mail: \email{alexand@ucy.ac.cy}}
\author{Martha Constantinou\\
       Department of Physics, University of Cyprus, P.O. Box 20537, 1678 Nicosia, Cyprus\\
        E-mail: \email{constantinou.martha@ucy.ac.cy}}
\author{Kyriakos Hadjiyiannakou\\
       Department of Physics, University of Cyprus, P.O. Box 20537, 1678 Nicosia, Cyprus\\
        E-mail: \email{hadjigiannakou.kyriakos@ucy.ac.cy}}
\author{Karl Jansen\\
       NIC, DESY, Platanenallee 6, D-15738 Zeuthen, Germany\\
        E-mail: \email{karl.jansen@desy.de}}
\author{Christos Kallidonis\\
 Computation-based Science and Technology Research  
    Center, Cyprus Institute, 20 Kavafi Str., Nicosia 2121, Cyprus\\
        E-mail: \email{c.kallidonis@cyi.ac.cy}}
\author{Giannis Koutsou\\
 Computation-based Science and Technology Research  
 Center, Cyprus Institute, 20 Kavafi Str., Nicosia 2121, Cyprus\\
        E-mail: \email{g.koutsou@cyi.ac.cy}}

\abstract{
We present results on the nucleon scalar, axial and tensor charges, as well as, 
on the first moments of the  unpolarized, polarized and transversity  parton distributions using  $N_f=2$ and $N_f=2+1+1$ twisted mass fermions. These  include an ensemble that yields the physical value of the ratio of the nucleon to the pion mass.
Results on the axial charges
of hyperons and charmed baryons are also presented for a range of pion masses
including the physical one. }

\FullConference{32nd International Symposium on Lattice Field Theory - LATTICE 2014\\
	  23-28 June, 2014\\
	Columbia University New York, NY}

\begin{document}

\vspace*{-0.6cm}

\section{Introduction}

\vspace*{-0.3cm}

We evaluate fundamental properties of the nucleon within the
fermion twisted mass formulation of lattice QCD.
The scalar, axial and tensor charges are computed using a number of twisted mass fermion (TMF) ensembles including one simulated with light quark masses fixed to their physical value.  The computation of the nucleon axial charge  allows a direct comparison with experiment, while 
determining the values of the scalar and tensor couplings provide useful input for searches 
beyond the familiar weak interactions of the Standard Model sought  in the decay of ultra-cold neutrons.
The computation of the tensor
charge is particularly timely since new experiments 
using polarized $^3$He/Proton  
at Jefferson lab aim at increasing the experimental accuracy by an order of magnitude~\cite{Gao:2010av}. In this works we use $N_f=2$ and $N_f=2+1+1$ ensembles of TMF
with lattice spacings smaller than 0.1~fm~\cite{Boucaud:2007uk} 
 to compute  nucleon matrix elements at zero momentum transfer.  The twisted mass formulation
is well-suited for hadron structure calculations since it provides automatic ${\cal O}(a^2)$ improvement requiring no operator modification~\cite{Frezzotti:2003ni}. Our results include a simulation using the Iwasaki gluon action, and $N_f=2$ TMF with a clover term yielding approximately the physical value of the pion mass, on a lattice of size $48^3\times 96$, referred to as the physical ensemble~\cite{Abdel-Rehim:2013yaa}. 

Using similar techniques as those for the case of the nucleon, 
we also compute the axial charges of hyperons and charmed baryons on the
same ensembles. The values of the  axial charges of these particles are 
either poorly measured or not known and thus lattice QCD  provides
valuable input  needed for phenomenological models. 

\vspace*{-0.3cm}

\section{Hadron spectrum}

\vspace*{-0.3cm}

\begin{figure}[h!]
\begin{minipage}{0.33\linewidth}
\hspace*{-0.3cm}\includegraphics[width=1.1\linewidth]{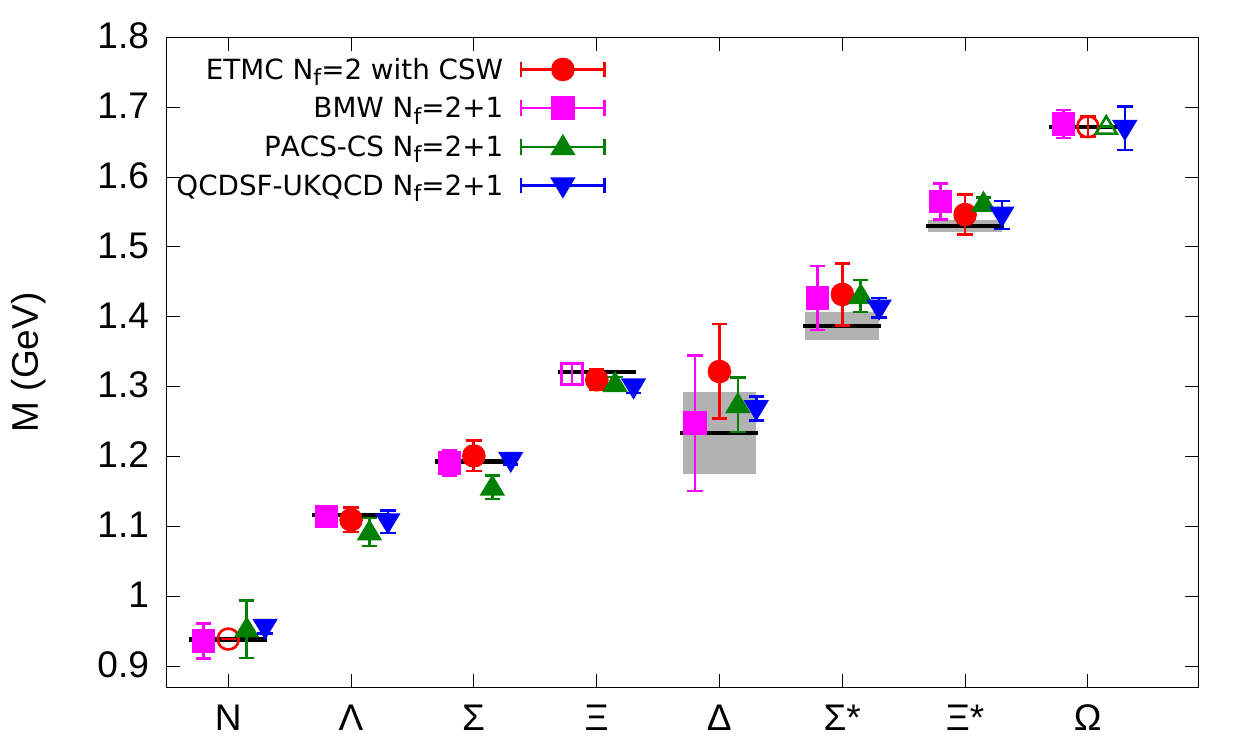}
\end{minipage}\hfill
\begin{minipage}{0.33\linewidth}
\hspace*{-0.15cm}\includegraphics[width=1.1\linewidth]{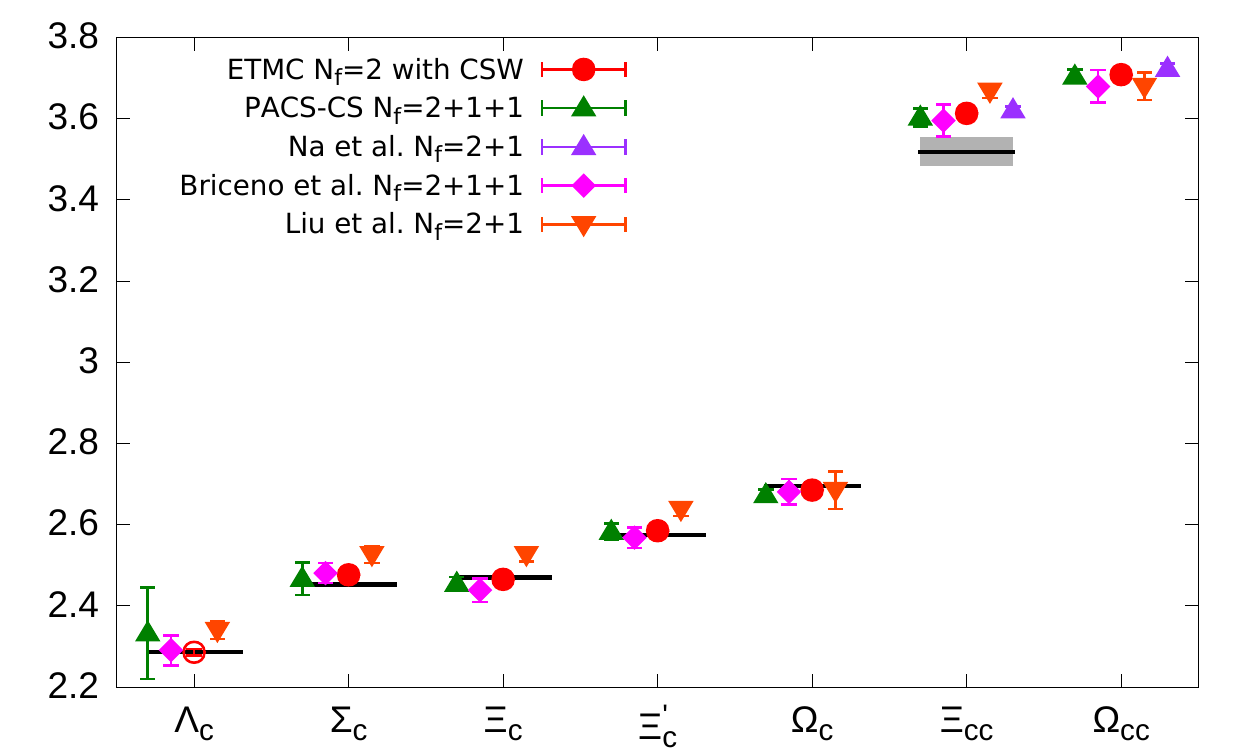}
\end{minipage}\hfill
\begin{minipage}{0.33\linewidth}
\includegraphics[width=1.1\linewidth]{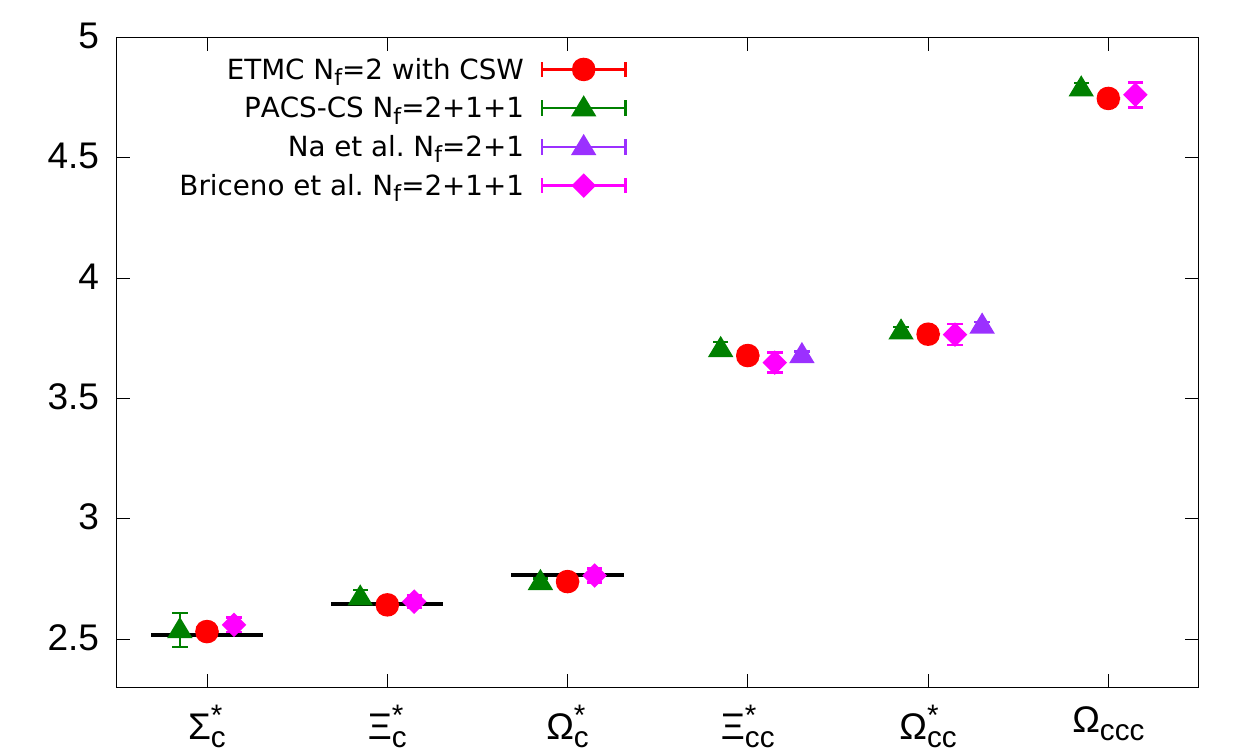}
\end{minipage}
\caption{Results on baryon masses using the physical ensemble. Left:  The octet and decuplet (the $\Omega$ and $\Xi$ were discovered at
Brookhaven 50 years ago~\cite{Barnes:1964ga}). Spin-1/2 (middle) and spin-3/2 (right) charmed baryons.}
\label{fig:spectrum}
\end{figure}

Before discussing the structure of baryons we need to compute their mass. 
In Refs.~\cite{Alexandrou:2014sha,Kallidonis}  the masses of baryons were
investigated using $N_f=2$ and $N_f=2+1+1$ TMF ensembles for three lattice spacings and a range of pion masses the smallest one being  210~MeV. Here, we 
present results computed for  the physical ensemble. The lattice spacing is
 $a=0.094(1)$~fm determined from the nucleon mass.
The strange and charm quark mass are fixed by using the $\Omega$ and $\Lambda_c$   mass, respectively. 
The resulting masses for the strange and charmed baryons  are shown in Fig.~\ref{fig:spectrum}. Although  the continuum limit is not yet performed   we observe agreement with the experimental values, as well as, with the values of other collaborations. Our  previous study suggests that
 discretization errors are small and this may explain agreement with experiment even at a finite value of $a$. Assuming negligible cut-off effects, we can give a values
 for the yet non-measured  mass of the charmed baryons. We find for the
mass of  $\Xi_{cc}^*$ 3.678(8)~GeV, for the  $\Omega^+_{cc}$ 3.708(10)~GeV, for $\Omega^{*+}_{cc}$  3.767(11)~GeV and for  $\Omega^{++}_{ccc}$ 4.746(3)~GeV.

\vspace*{-0.5cm}

\section{Nucleon structure}

\vspace*{-0.3cm}

Results on the nucleon axial $g_A$, scalar $g_S$  and tensor $g_T$  charges and three first
moments  $\langle x\rangle_{q}$, $\langle x\rangle_{\Delta q}$ and  $\langle x\rangle_{\delta q}$  are extracted by computing the forward nucleon  matrix element $\langle N(\vec{p^\prime}){\cal O}_X^a N(\vec{p})\rangle|_{q^2=0}$. The operator
${\cal O}_{X}^a$ is $\bar{\psi}(x)\frac{\tau^a}{2}\psi(x)$ for the scalar,
$\bar{\psi}(x)\gamma^{\mu}\gamma_5\frac{\tau^a}{2}\psi(x)$ for the axial-vector,  and
$\bar{\psi}(x)\sigma^{\mu\nu}\frac{\tau^a}{2}\psi(x)$ for the tensor charges, with the corresponding expressions for one derivative operators for  the first moments. Here we present results for two ensembles:
i)  $N_f=2+1+1$ TMF on a $32^3\times 64$ lattice with $a$=0.082~fm and $m_\pi=373$~MeV (referred to as B55). A high statistics analysis is performed including disconnected contributions, using seven sink-source time separations ranging from 0.5~fm to 1.5~fm, and ii)
$N_f=2$  twisted mass with  clover term on a $48^3\times 96$ lattice with  $a$ = 0.094(1)~fm and $m_\pi$ = 130~MeV (physical ensemble). A total of  $\sim 1000$ confs are analyzed for three sink-source time separations ranging from about 0.9~fm to 1.3~fm.  

\begin{figure}[h!]\vspace*{-0.3cm}
\begin{minipage}{0.49\linewidth}
\hspace*{-0.2cm}\includegraphics[width=1.05\linewidth]{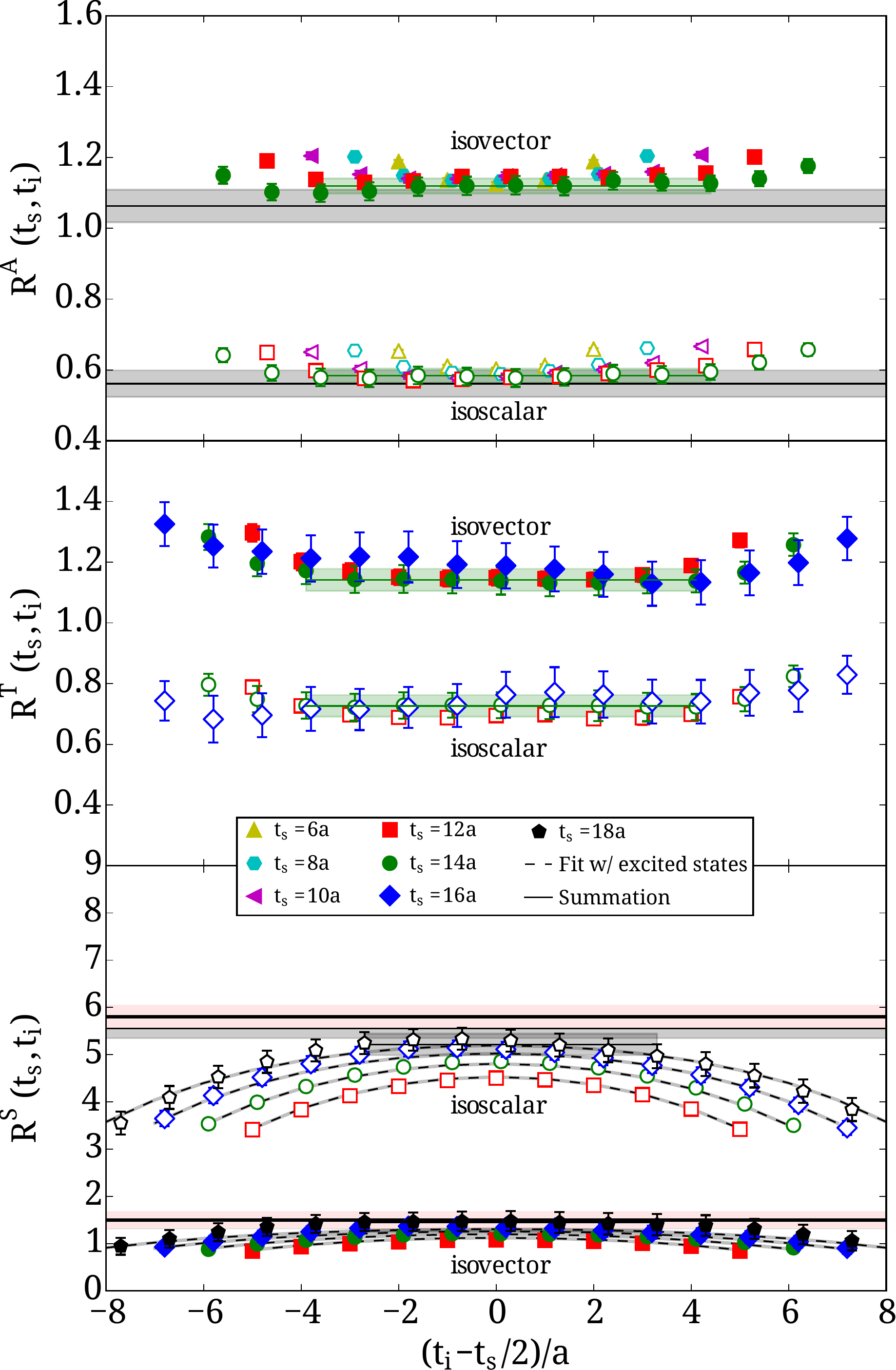}
  \end{minipage}\hfill
 \begin{minipage}{0.49\linewidth}
\includegraphics[width=1.05\linewidth]{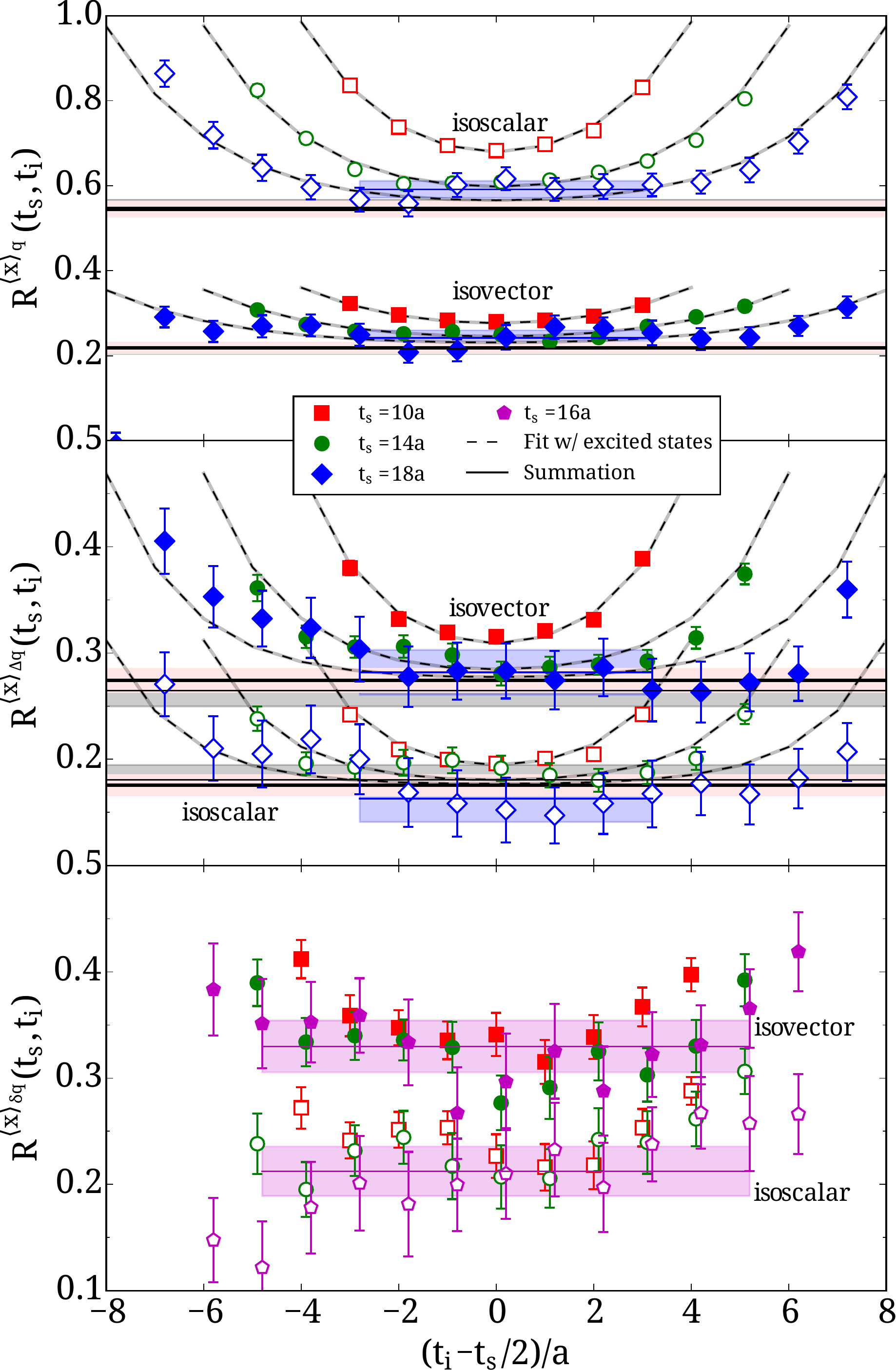}
\end{minipage}
\caption{Results on the connected ratio  $R(t_s,t_{\rm ins},t_0)$ for B55  using  1200 statistics for various $t_s$.  Left panel:  axial (top), tensor (middle) and scalar (bottom) currents. Right panel:  one derivative vector, axial and tensor operators.
The grey band is the result from the summation method. We also show simultaneous fits that take into account one excited state (dashed lines) yielding the value for the matrix element shown in light red. }  
\label{fig:B55}
\vspace*{-0.3cm}
\end{figure}

\begin{figure}[h!] 
 \begin{minipage}{0.49\linewidth}
\includegraphics[width=\linewidth]{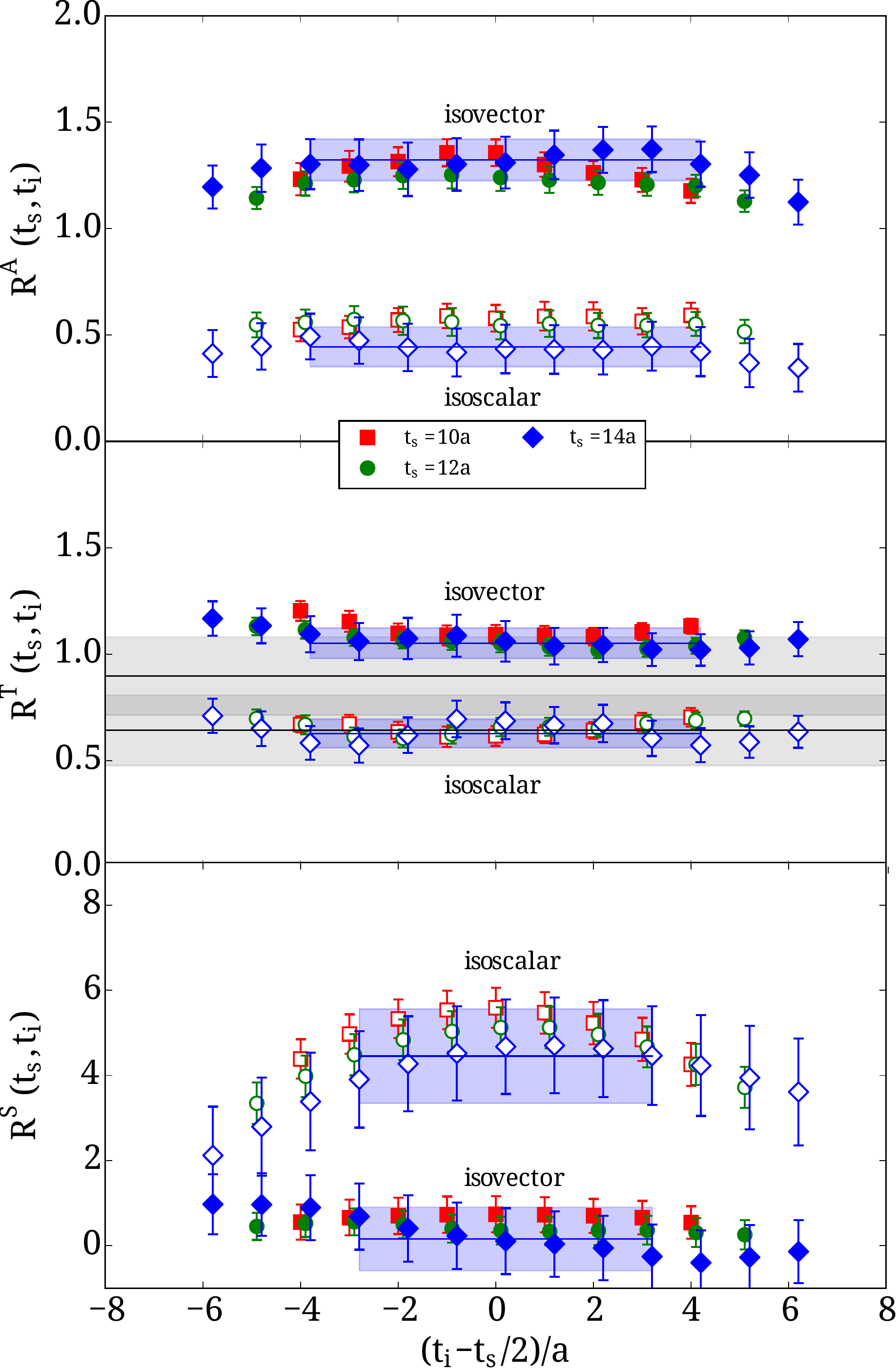}
  \end{minipage}\hfill
\begin{minipage}{0.49\linewidth}
\includegraphics[width=\linewidth]{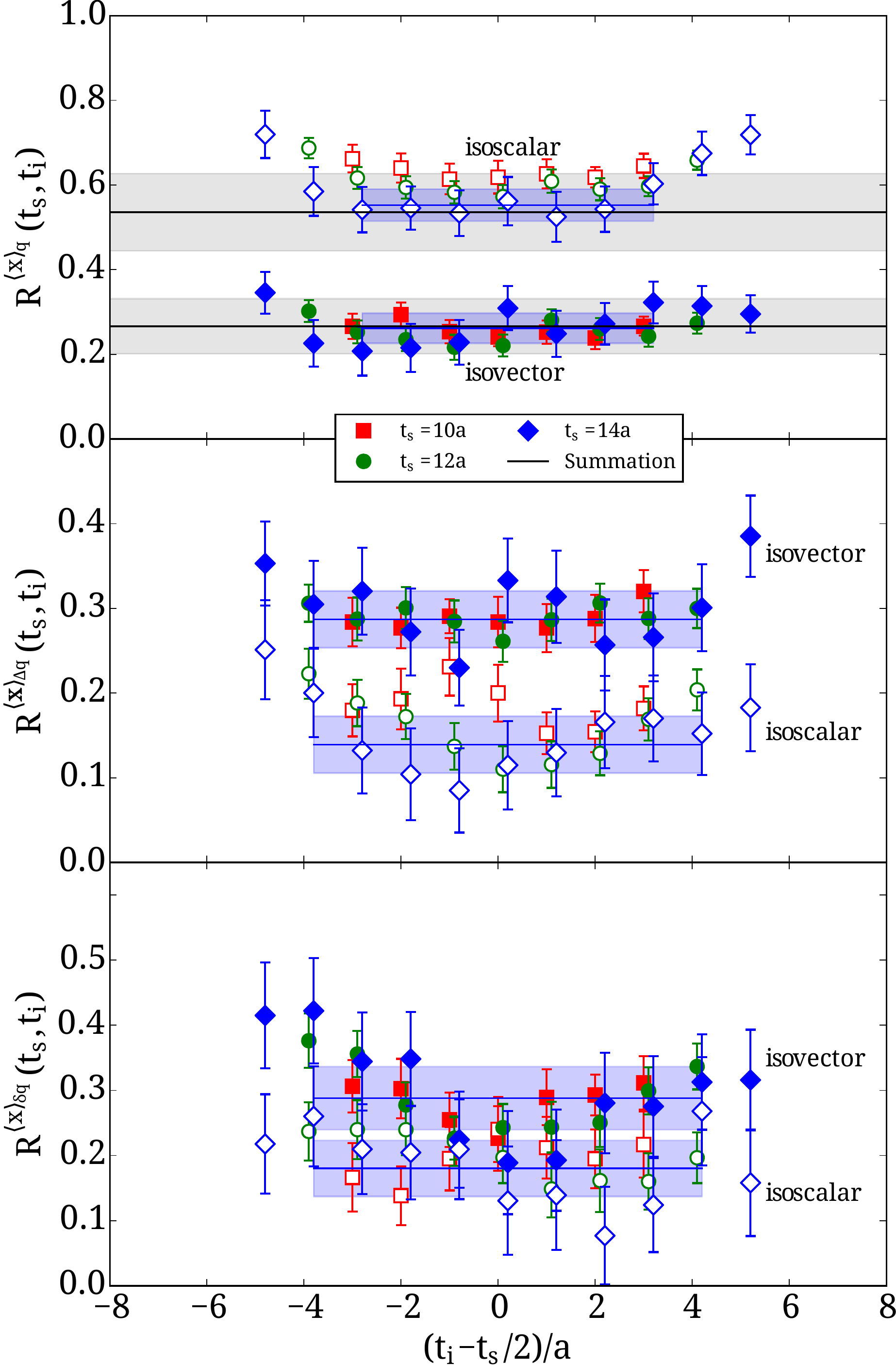}
  \end{minipage}
\caption{Results for the connected ratio $R(t_s,t_{\rm ins},t_0)$   for the physical ensemble for several $t_s$ using ${\cal O}(10^3)$ statistics. The notation is the same
as in Fig.~\protect\ref{fig:B55}.}
\label{fig:physical}
\vspace*{-0.4cm}
\end{figure}

The nucleon matrix element is extracted from the
ratio $ R(t_s,t_{\rm ins},t_0) $ of the relevant three-point correlator divided by an appropriate combination of two-point functions~\cite{Alexandrou:2013joa} after identifying
the nucleon ground state. Various approaches are applied to check ground state dominance:
i) Plateau method:\vspace*{-0.1cm}
 \begin{equation}
    R(t_s,t_{\rm ins},t_0) \xrightarrow[(t_s-t_{\rm ins})\Delta \gg 1]{(t_{\rm ins}-t_0)\Delta \gg 1} \mathcal{M}[1
      + \dots e^{-\Delta({\bf p})(t_{\rm ins}-t_0)} + \dots e^{-\Delta({\bf
          p'})(t_s-t_{\rm ins})}]\nonumber
  \end{equation}
where $\mathcal{M}$ is the desired matrix element,  $t_s,t_{\rm ins},t_0$ are the
  sink, insertion and source time-slices
and $\Delta({\bf p})$ the
  energy gap of the first excited state; 
ii) Summation method defined by  summing over $t_{\rm ins}$:
 $
 \sum_{t_{\rm ins}=t_0-1}^{t_s-1} R(t_s,t_{\rm ins},t_0) = {\sf Const.} + \mathcal{M}[(t_s-t_0) + \mathcal{O}(e^{-\Delta({\bf p})(t_s-t_0)})  + \mathcal{O}(e^{-\Delta({\bf p'})(t_s-t_0)})].
  $
In this approach excited state contributions are suppressed by exponentials decaying with $t_s-t_0$, rather than $t_s-t_{\rm ins}$ and $t_{\rm ins}-t_0$.
 However, one needs to fit the slope rather than to a constant or take differences and then fit to a constant~\cite{Maiani:1987by, Capitani:2012gj}. Both procedures  generally lead to larger errors; 
iii) Perform fits including the contribution due to the first excited state.

Let us first  examine the 
isovector combination, which has no disconnected contributions.
In Fig.~\ref{fig:B55}  we show the ratios from which
$g_A$, $g_T$ and $g_s$ are extracted. 
As can be seen from analyzing a number of sink-source separations for the
B55 ensemble,  $g_A$ shows no detectable excited states and  $g_T$ behaves similar to $g_A$~\cite{Alexandrou:2013wka}, while $g_s$ has severe contamination from excited states.
The case of $g_s$ illustrates nicely that the plateau, summation and two-state fits all give consistent results. When this happens we have confidence in ground state dominance  extracting a value that corresponds to the desired matrix element~\cite{Alexandrou:2013jsa}.
A similar analysis is performed for the 
first moments: $
\langle x \rangle_q=\int_0^1 dx \> x\left[q(x)+\bar{q}(x)\right],\,\,
\langle x\rangle_{\Delta q}=\int_0^1 dx \> x\left[\Delta q(x)-\Delta \bar{q}(x)\right]$,
 and
$\langle x\rangle_{\delta q}=\int_0^1dx \>x\left[\delta q(x)+\delta \bar{q}(x)\right]$,
where
 $q(x)=q(x)_\downarrow+q(x)_\uparrow\,,\Delta q(x)=q(x)_\downarrow-q(x)_\uparrow\,, \delta q(x)=q(x)_\perp+q(x)_\top$
for the one derivative vector, axial-vector and tensor operators respectively.
 Results on the three lowest moments obtained in the $\overline{\rm MS}$ at 4~GeV$^2$ are shown in Fig.~\ref{fig:B55} for  B55. As can be seen, there is  noticeable excited state contamination, especially for the isoscalar. Again consistency of the results extracted using the plateau and summation methods,  and the two-state fit gives confidence in the final value.
 In Fig.~\ref{fig:physical} we show the corresponding analysis for the physical ensemble. Although the statistics is similar to that for B55 i.e. ${\cal O}(10^3)$ the errors are larger as expected, requiring more statistics to extract final numbers. Thus our results for the physical ensemble are to be regarded as preliminary.

\begin{figure}[h!]
\begin{minipage}{0.49\linewidth}
\hspace*{-0.5cm}\includegraphics[width=1.1\linewidth,height=1.6\linewidth]{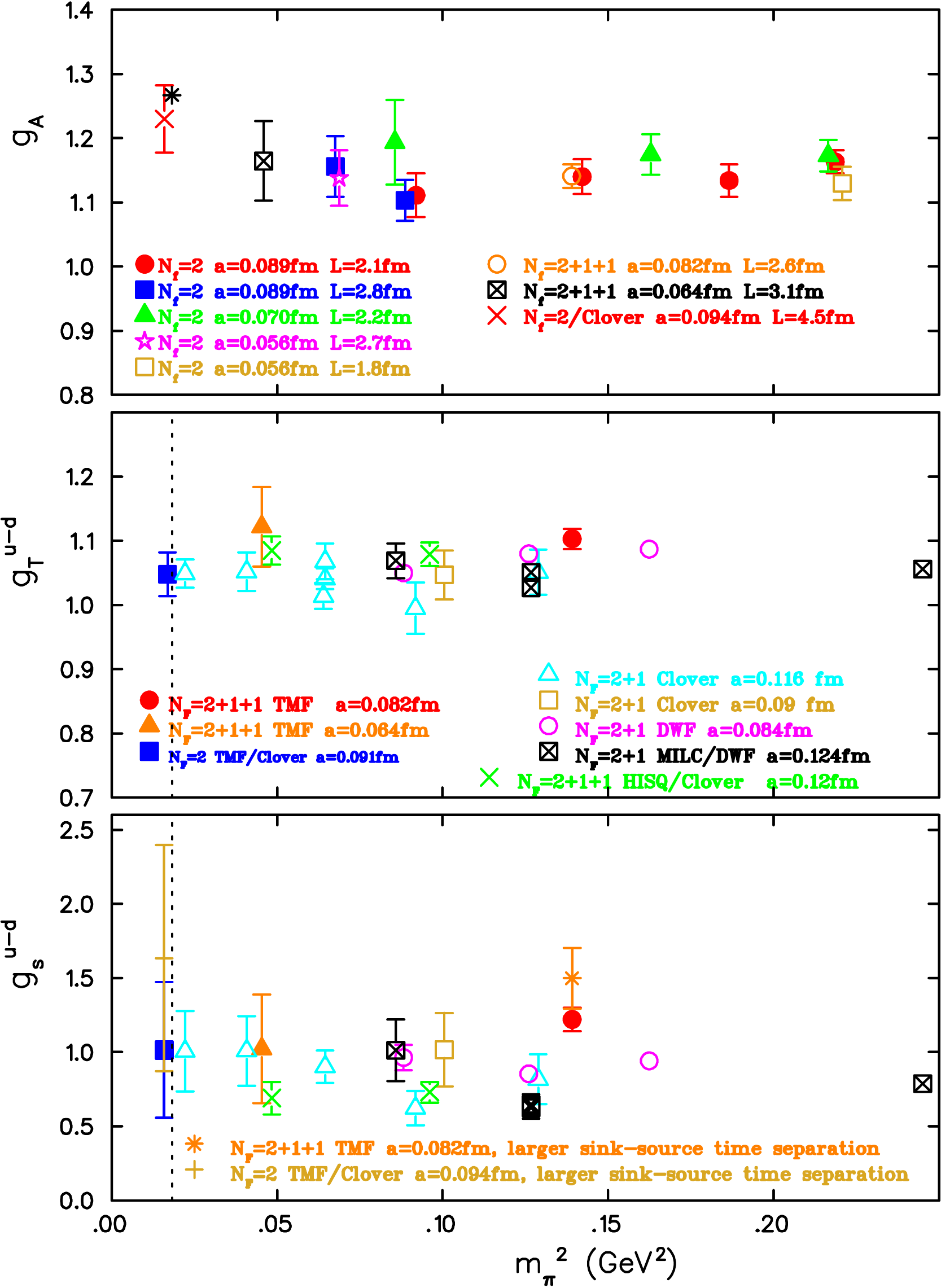}
\end{minipage}\hfill
\begin{minipage}{0.49\linewidth}
\includegraphics[width=1.1\linewidth,height=1.6\linewidth]{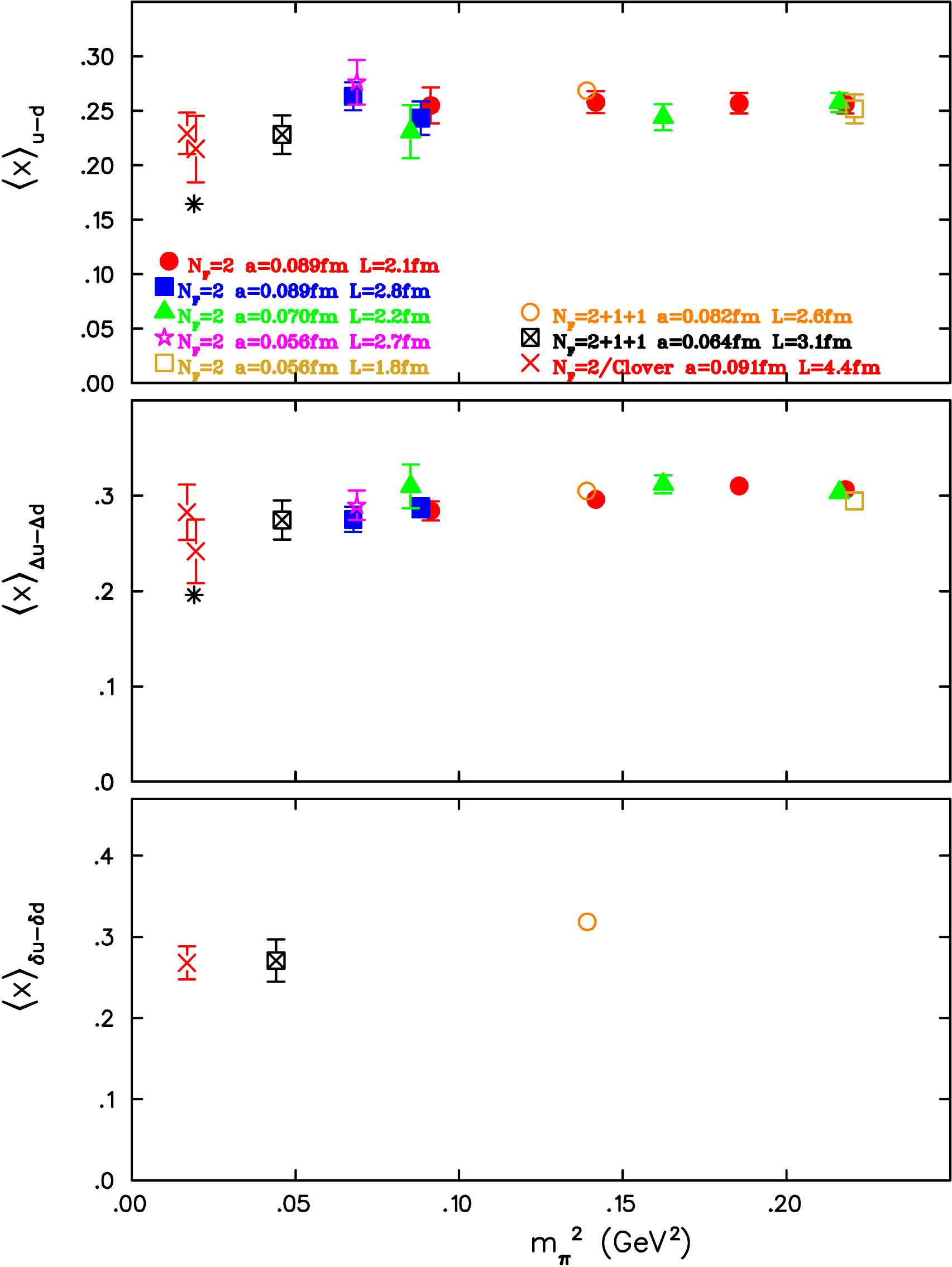}
\end{minipage}
\caption{Nucleon isovector charges  $g_A$, $g_T$ and $g_s$ (left) and  first moments  $\langle x \rangle_q$, $\langle x \rangle_{\Delta q}$ and 
$\langle x \rangle_{\delta q}$ 
(right) versus $m_\pi^2$. TMF results at the physical point are preliminary. 
For $g_T$ and $g_s$ we include a comparison with the results of other collaborations.
The orange asterisk and cross for $g_s$ at $m_\pi=373$~MeV and 130~MeV respectively, is for $t_s-t_0\sim 1.5$~fm, while the rest of the data are for $t_s-t_0\sim 1.1-1.2$~fm. }
\label{fig:results}
\end{figure}

In order to evaluate the isoscalar nucleon  charges $g^{u+d}_s$, $g^{u+d}_A$ and $g^{u+d}_T$ one needs the disconnected contributions. For B55 these were computed for all local and one derivative operators using about 150,000 measurements on GPUs~\cite{Alexandrou:2013wca,Abdel-Rehim:2013wlz,Alexandrou:2014fva,Alexandrou:2014eva,Alexandrou:2014}. We found that  disconnected contributions to  $g^{u+d}_A$ are of the order of 10\% while for $g^{u+d}_T$ are consistent with zero.The disconnected contributions  to the  $g^{u+d}_s$ arising from the light quark loops  are larger than 10\%, while from the strange are smaller by a factor of about three. From this study, it is clear that we need to include the disconnected contributions in order to extract reliable results on  $g^{u+d}_A$ and $g^{u+d}_s$.

In Fig.~\ref{fig:results} we collect our results for the nucleon isovector charges and first moments for $t_s-t_0\sim 1.0-1.2$~fm, which may be  sufficient for quantities like $g_A$ 
but not for others such as  $g_s$ where excited state contamination requires $t_s-t_0>1.5$~fm. Indeed the agreement of $g_A$ at the physical pion mass with the experimental  value indicates that such lattice artifacts are small. Given that  $g_T$ shows similar behavior to $g_A$ the value found at the physical point can be taken as a prediction for the tensor charge of the nucleon. 
 For $g_s$ one needs to  increase the sink-source time separation to $\sim 1.5$~fm to ensure ground state dominance. Therefore, the results shown in Fig.~\ref{fig:results} using $t_s-t_0\sim 1.1$~fm are only preliminary and we need
to study  $g_s$  as we increase $t_s-t_0$.  
Similarly  $\langle x\rangle_{u-d}$ and  $\langle x\rangle_{\Delta u-\Delta d}$ approach the physical value for bigger sink-source separations. In order
to finalize their value one would need an equivalent high statistics study as the one performed at  $m_\pi=373$~MeV.

\vspace*{-0.3cm}

\section{Axial charges of hyperons and charmed baryons}

\vspace*{-0.3cm}

\begin{figure}[h!]
\begin{minipage}{0.49\linewidth}
\includegraphics[width=\linewidth]{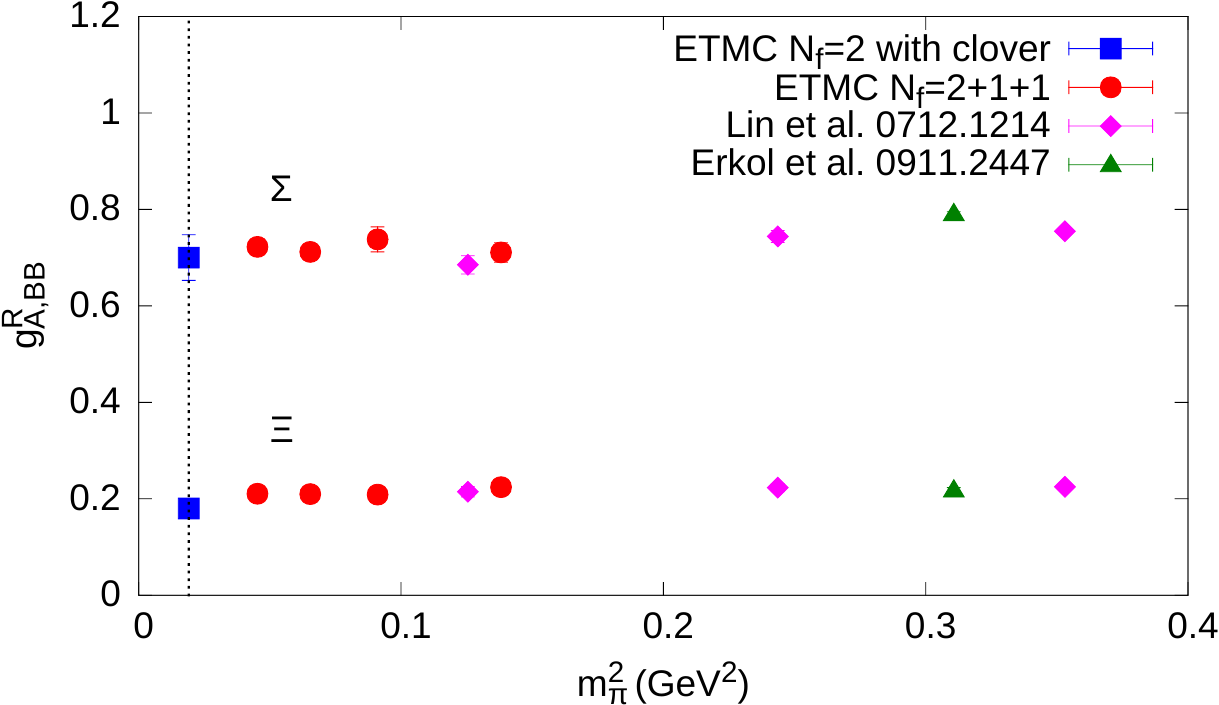}
\end{minipage}
\begin{minipage}{0.49\linewidth}
\hfill
\includegraphics[width=\linewidth]{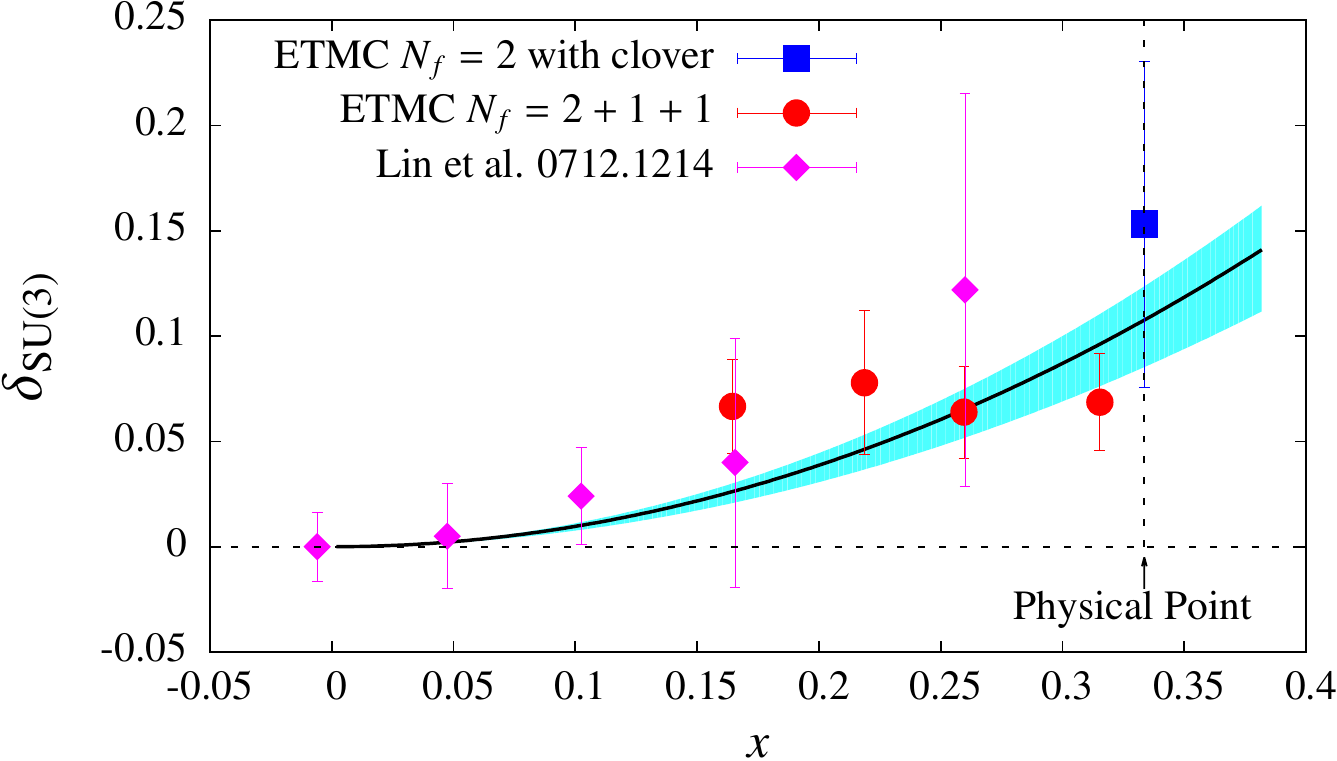}\\
\end{minipage}
\caption{Left: Axial charge for the $\Sigma$ and $\Xi$ baryons versus $m_\pi^2$. Right: SU(3) breaking $\delta_{SU(3)}=g_A^N-g_A^\Sigma+g_A^\Xi$ versus $x=\left(m_K^2-m_\pi^2\right)/(4\pi^2f_\pi^2)$.}
\label{fig:SU3 breaking}
\vspace*{-0.5cm}
\end{figure}

\begin{figure}[h!]
\begin{minipage}{0.49\linewidth}
\includegraphics[width=\linewidth]{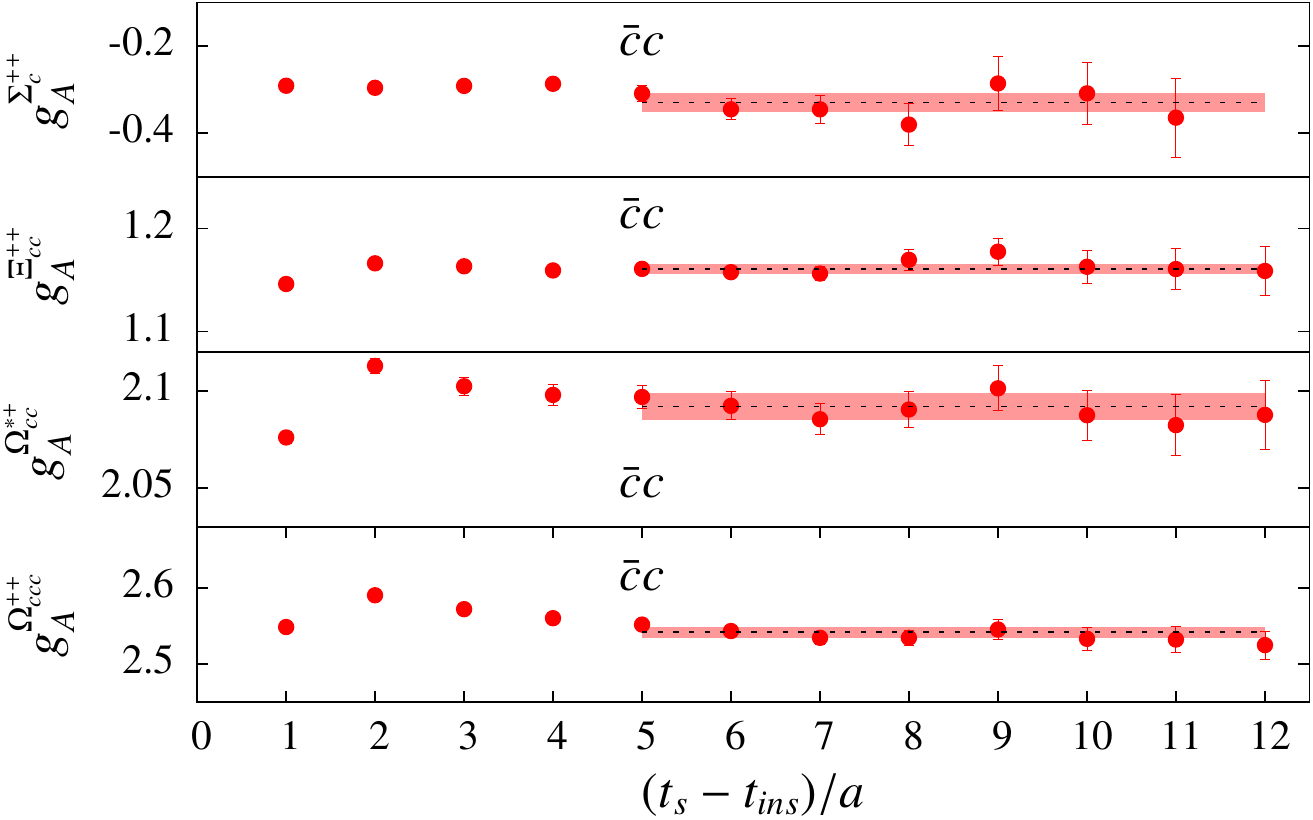}
\end{minipage}
\begin{minipage}{0.49\linewidth}
\hfill
\includegraphics[width=\linewidth]{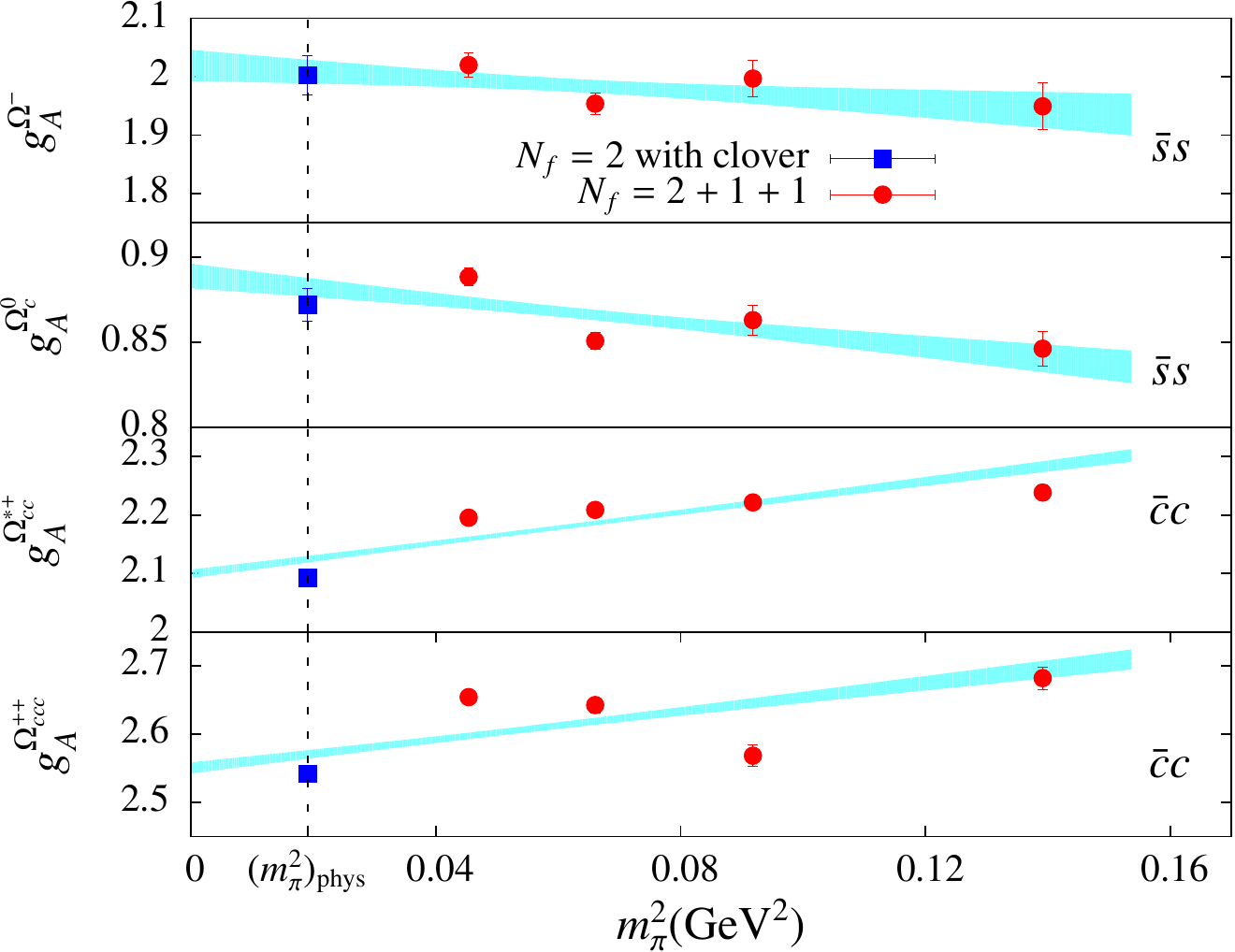}
\end{minipage}
\caption{Left: Representative ratios from which the charmed contribution to the axial charge is extracted versus the sink time for fixed $t_{\rm ins}$. Right: Axial charges of the $\Omega$ baryons versus $m_\pi^2$. }
\label{fig:charm gA}
\vspace*{-0.2cm}
\end{figure}

 Having the formalism to compute the nucleon  axial charge one can use similar techniques to extract the axial charges of hyperons and charmed baryons. For this calculation we consider only connected contributions and  use the fixed current method to evaluate the  
axial matrix elements $\langle B(\vec{p}^\prime)|\bar{\psi}(x)\gamma_\mu\gamma_5\psi(x)|B(\vec{p})\rangle|_{q^2=0}$  performing only one
sequential inversion per quark flavor for all baryons.
In Fig.~\ref{fig:SU3 breaking} we show representative results for the pion mass dependence of the axial charges of the $\Sigma$ and the $\Xi$ baryons. As can be seen,
the dependence  in $m_\pi^2$ is rather weak.
In the same figure we also show the SU(3) breaking parameter $\delta_{\rm SU(3)}\equiv g_A^N-g_A^\Sigma+g_A^\Xi$.
As one moves away from the $SU(3)$-symmetric point $\delta_{\rm SU(3)}$ increases reaching about 0.15 at the physical point. 
In Fig.~\ref{fig:charm gA} we show results on the axial charges of charmed baryons. The ratio $R(t_s,T_{\rm ins},t_0)$ at fixed $t_{\rm ins}/a=7$ versus $t_s$ shows nice plateaus allowing a good determination of the axial charges.
The axial charges of the $\Omega$ baryons show a linear dependence in $m_\pi^2$ yielding a result in complete agreement with the direct determination using our physical ensemble which is not included in the fit.

\vspace*{-0.3cm}

\section{Conclusions}

\vspace*{-0.3cm}

 Simulations at the physical point are now feasible and this opens exciting 
possibilities for the study of hadron structure. In this work we show 
results on  the nucleons charges, as well as, on the three first moments of PDFs.
We find a value of $g_A$ that is in agreement with experiment.
However, our results at the physical point highlight the need for larger statistics in order to carry out   careful cross-checks.
Noise reduction techniques  such as  all-mode-averaging, improved methods for disconnected diagrams and smearing techniques are currently being pursued aiming at  decreasing our errors on the quantities obtained at the physical point.
 Predictions for other hadron observables are  emerging as shown here for 
the case of the  axial
charges of hyperons and charmed baryons, as well as, on quantities  probing beyond the standard model physics such as  $g_T$ and $g_s$ from which the closely related $\sigma$-terms can  be also extracted.

\vspace*{0.3cm}

\noindent
{\bf Acknowledgements:}
Partial support was provided by the projects EPYAN/0506/08,
TECHNOLOGY/ $\Theta$E$\Pi$I$\Sigma$/0311(BE)/16, $\Pi$E/$\Pi\Lambda$HPO/0311(BIE)/09 and 
$\Pi$PO$\Sigma$E$\Lambda$KY$\Sigma$H/EM$\Pi$EIPO$\Sigma$/0311/16 funded by the Cyprus Research Promotion Foundation.
 This work used computational resources provided by PRACE, JSC, Germany and the  Cy-Tera project (NEA Y$\Pi$O$\Delta$OMH/$\Sigma$TPATH/0308/31).

\end{document}